\documentclass[fleqn,usenatbib]{mnras}

\usepackage{newtxtext,newtxmath}
\usepackage{ulem}
\usepackage{natbib}

\usepackage[T1]{fontenc}

\DeclareRobustCommand{\VAN}[3]{#2}
\let\VANthebibliography\thebibliography
\def\thebibliography{\DeclareRobustCommand{\VAN}[3]{##3}\VANthebibliography}

\usepackage{graphicx}	% Including figure files
\usepackage{amsmath}	% Advanced maths commands
\title[]{Searching for radio pulsation from SGR 1935+2154 with the Parkes Ultra-Wideband Low receiver}
\author[Zhenfan Tang et al.]{
Zhenfan Tang,$^{1,2}$
Songbo Zhang,$^{1}$ $^\ast$
Shi Dai,$^{3}$
Ye Li,$^{1}$
Xuefeng Wu$^{1}$ $^\ast$
\\
$^{1}$Purple Mountain Observatory, Chinese Academy of Sciences, Nanjing 210023, China\\
$^{2}$School of Astronomy and Space Sciences, University of Science and Technology of China, Hefei 230026, China\\
$^{3}$Western Sydney University, Locked Bag 1797, Penrith South DC, NSW 1797, Australia
}

\pubyear{2021}

\begin{document}
\label{firstpage}
\pagerange{\pageref{firstpage}--\pageref{lastpage}}
\maketitle

%%% p1-Abstract
\begin{abstract}
Magnetars have been proposed to be the origin of FRBs soon after its initial discovery. The detection of the first Galactic FRB 20200428 from SGR 1935+2154 has made this hypothesis more convincing. In October 2020, this source was supposed to be in an extremely active state again.
We then carried out a 1.6-hours follow-up observation of SGR 1935+2154 using the new ultra-wideband low (UWL) receiver of the Parkes 64\,m radio telescope covering a frequency range of 704$-$4032 MHz. However, no convincing signal was detected in either of our single pulse or periodicity searches. 
We obtained a limit on the flux density of periodic signal of $\rm 3.6\,\mu Jy$ using the full 3.3\,GHz bandwidth data sets, which is the strictest limit for that of SGR 1935+2154. Our full bandwidth limit on the single pulses fluence is 35mJy ms, which is well below the brightest single pulses detected by the FAST radio telescope just two before our observation. Assuming that SGR 1935+2154 is active during our observation, our results suggest that its radio bursts are either intrinsically  narrowband or show a steep spectrum.
\end{abstract}


\section{Introduction}

%%% FRB  review
%Fast radio bursts are expected to have large luminosity (up to $10^{39} {\rm erg\, s^{-1}}$) and one of the most energetic sources in the universe.

Fast radio bursts are one of the most energetic sources in the universe with luminosities up to $10^{39} {\rm erg\, s^{-1}}$.
%With the high peak energy in radio band (up to $10^{39} {\rm erg/s}$) , FRB is one of the most energetic sources in the universe.
Since the original discovery in 2007\citep{lorimerBrightMillisecondRadio2007}, efforts to explore the  physical origin of FRBs have continued. 
%a lot of efforts have been devoted to uncovering their physical origin. 
Several important progress has been made in the last few years, including the localization for host galaxy and detection of periodic activities~\citep{chatterjeeDirectLocalizationFast2017, thechime/frbcollaborationPeriodicActivityFast2020}.   
%A lot of progress have been made particularly in last few years, like the localization for host galaxy and finding of periodic activities  \citep{chatterjeeDirectLocalizationFast2017, thechime/frbcollaborationPeriodicActivityFast2020}. 
%
%200428
FRB~200428, a Galactic FRB event detected by the Canadian Hydrogen Intensity Mapping Experiment (CHIME) and the Survey for Transient Astronomical Radio Emission 2 (STARE2), is another breakthrough in revealing the mystery of FRB origin~\citep{Andersen:2020hvz, bochenekFastRadioBurst2020}.
%The detection for FRB 200428, a galatic FRB event observed by the Canadian Hydrogen Intensity Mapping Experiment (CHIME) and the Survey for Transient Astronomical Radio Emission 2 (STARE2) made another greet leap in revealing the mystery 
Considering the dispersion delay, the two X-ray components of the magnetar burst occur within 3 ms of the radio burst components \citep{2020ATel13696....1Z}.
%Considering the DM delay, the two components of the magnetar burst detected by several X-ray telescopes occur within 3 ms of the radio burst components.

%理论讨论
Magnetars have been proposed to be the origin of FRBs~\citep{popovHyperflaresSGRsEngine2007} soon after its initial discovery. A large number of papers discussed this model from different perspectives~\citep{2014MNRAS.442L...9L, katzHOWSOFTGAMMA2016}.  
The detection of FRB~20048 shows that magnetars are able to generate bright radio bursts with luminosity close to FRBs. However, extreme activities of some FRBs (e.g., FRB~121102 \citep{gajjarHighestFrequencyDetection2018}) are still not understood and most of the FRBs are much more energetic than FRB~200428.      
%This discovery shows magnetars are capable of producing FRB types of hight energy radio bursts.
%So the questions come next are how magnetar generate such dramatic radio outburst, and can magnetars explain all the FRB events.
%
%It had been proposed soon after the discovery of the Lorimer burst that magnetars can produce FRB \citep{popovHyperflaresSGRsEngine2007}.
%This theory has been discussed by many people\citep{2014MNRAS.442L...9L, katzHOWSOFTGAMMA2016}.
There are generally two types of coherent radio emission models, those originating in the magnetospheres and those produced by relativistic shocks \citep{zhangPhysicalMechanismsFast2020}.
%Put aside the details, there are generally two types of coherent radio emission models, those invoking the magnetospheres and those invoking relativistic shocks \citep{zhangPhysicalMechanismsFast2020}.
%
Such models can explain the energy ratio of FRB~200428 and its associated X-ray burst (XRB), but the magnetosphere origin has already been well established to explain the XRBs of magnetars and are currently the most promising models for FRB~20048-like events.
%They all can explain the X-ray radio ratio for FRB 200428 and the associated X-ray burst (XRB).
%But the XRBs from magnetars are already well established with the magnetosphere origin \citep{thompson1995soft}.
%So, for now magnetosphere models are more favorited.

%%% magnetar
Magnetars are a small group of neutron stars with long rotation periods and high slow-down rates, which indicates an extremely high surface magnetic field (> $10^{14} \rm G$)
\citep{kaspiMagnetars2017}. 
More than 30 magnetars have been discovered so far~\footnote{http://www.physics.mcgill.ca/~pulsar/magnetar/main.html}.
%There are now 30 known magnetars \footnote{http://www.physics.mcgill.ca/~pulsar/magnetar/main.html}.
Most of them were discovered by X-ray observations thanks to their wide range of X-ray activity, including short bursts, large outbursts, and giant flares.   
%They are mostly found by X-ray observation due to their wide range of X-ray activity including short bursts, large out-bursts and giant flares.
The quasi-periodic oscillations in the tails of their giant flares and associations with supernova remnants prove their neutron-star origin \citep{mazets1979observations, 1982ApJ...255L..45C}.
X-ray luminosities of magnetars are much larger than their rotational energy loss, and therefore their emission and bursts are widely believed to be powered by large magnetic fields.
%These sources are widely believed to be powered by the decay of their large magnetic fields, because the X-ray luminosity are far beyond their rotational energy loss. 

%射电辐射
Only six magnetars have shown radio pulsations.  
%There are only six magnetars have been detected with radio pulsations.
Their radio pulsations were mostly detected during the decay of X-ray emission. 
\citep{camilo2007variable}.
Spectra of these radio emissions are remarkably flat, different from the normal pulsar population whose spectra are steep with negative spectral indices of $\sim -1.8 \rm$ \citep{maron2000pulsar},
except for one magnetar SGR 1745–2900~\citep{pennucciSIMULTANEOUSMULTIBANDRADIO2015}.
%But there is one exception for SGR 1745–2900 \citep{pennucciSIMULTANEOUSMULTIBANDRADIO2015}.
%
Bright radio single pulses of magnetars are similar to giant pulses(GPs) of pulsars, with a power-law fluence distribution and shorter duration than average pulsation profile \citep{espositoVeryYoungRadioloud2020}.

%%% 观测回顾

% sgr1935的发现和认知
SGR J1935+2154 was discovered by Swift-BAT in 2014 through its magnetar-like bursts\citep{2014GCN.16520....1S} and cemented by the following Chandra and XMM-Newton observations~\citep{israelDiscoveryMonitoringEnvironment2016}. 
%The following Chandra and XMM-Newton observation cemented its identification \citep{israelDiscoveryMonitoringEnvironment2016}.
Its spin period and time derivative of the period are 3.24 s and $1.43\times10^{-11} {\rm s\,s}^{-1}$,    
%It spin at a period of about 3.24 s, and slow down at a rate of $1.43\times10^{-11} s s^{-1}$.
which implies a surface dipolar magnetic field strength of $~2.2\times10^{14}$ G, and a characteristic age of about 3.6 kyr.
These properties make SGR~J1935+2154 a typical Galactic magnetar.
%These propert make SGR J1935+2154 to be one typical of the Milky Way magnetar population.
%
Its position strongly suggesting an association with a supernova remnant (SNR) G57.2+0.8 at a distance of $\rm \sim9 kpc$.
\citep{2014GCN.16533....1G, zhou2020revisiting}
Observations of several radio telescopes  failed to detect any pulsed or persistent radio emission after the discovery of SGR J1935+2154, and no pulsar wind nebula(PWN) has been found~\citep{2014GCN.16542....1F, 2014ATel.6376....1S, 2014ATel.6371....1B}
%Searches with the multiple radio telescope failed to detect any pulsed or persistent radio emission after the fist finding
%\citep{2014GCN.16542....1F, 2014ATel.6376....1S, 2014ATel.6371....1B}
%And so far no pulsar wind nebula(PWN) was found.
In 2015, 2016 and 2019 this source entered active state and showed burst activities more frequently and intensely~\citep{2017ApJ...847...85Y, linFermiGBMView2020}.
Even during the quiescent time, several sporadic XRBs have been detected, which makes it outstanding upon other known magnetars~\citep{2017ApJ...847...85Y}.   
%But even in the time between these active period, it also shown sporadic XRBs, which makes it outstanding upon known magnetars
%\citep{2017ApJ...847...85Y}.

% 4月
On April 27, 2020, multiple X-ray bursts were detected from SGR J1935+2154, indicated a new active  phase \citep{2020GCN.27657....1B}.
One day later, FRB 200428 was detected associated with two SGR bursts \citep{2020ATel13696....1Z}.
After its outburst in April, a number of radio telescopes have undertaken follow-up observations of SGR~1935+2154. Only afew radio bursts were detected~\citep{2020ATel13699....1Z, 2021NatAs...5..414K}.   
X-ray observations showed that the black body temperature and unabsorbed flux in the 0.3-10\,keV band of this magnetar have gone through a double exponential decay, and went back to average values three months later~\citep{younesNICERView20202020}.

% 10月
On October 8, 2020, CHIME detected three close bursts with fluence of  $900 \pm 160$, $9.2 \pm 1.6$ and $6.4 \pm 1.1$ Jy ms, respectively~\citep{2020ATel14074....1G, 2020ATel14080....1P}. A XRB of SGR~1935+2154 was reported by Swift soon after, but was later to be a detector glitch~\citep{2020ATel14076....1T}.
One day later, during a one-hour observation, FAST detect multiple radio pulses with fluence up to 40 mJy ms
\citep{2020ATel14084....1Z}.
They also detected a periodic signal with a period of 3.24781s.
And single pulses were well aligned in a certain phase of the period.

% 章节介绍
We have also carried out a follow-up observing campaigns using Parkes after the outburst. Here we report the details of this observation and our results.
The observation and data reduction are described in Section~\ref{observation}. The results are presented in Section~\ref{res} and we discuss the possible implications from our observation in Section \ref{dis}.

%%%%% p3-Observation 
\section{Observation and Data Reduction} \label{observation}

%%%\subsection{Observation}
During the reactivation of SGR~1935+2154 in October 2020, we carried out an 1.6hr follow-up observation with the Parkes 64\,m radio telescope on October 11, 2020. We used the new ultra-wideband low (UWL) receiver system~\citep{hobbsUltrawideBandwidth7042020} covering a frequency range of 704$-$4032\,MHz.The full band is split into 26 contiguous sub-bands, each with 128 channels. The channelised signals were recorded with all four polarisations using Parkes Medusa digital systems and 8-bit sampled data with a resolution of 64$\mu$\,s to be stored in PSRFITS search mode format \citep{hotanPSRCHIVEPSRFITSOpen2004}.
As the reported DM of SGR~1935+2154 is around 333 pc $cm^{-3}$ \citep{thechime/frbcollaborationBrightMilliseconddurationRadio2020}, were coherently de-dispersed the data at a DM of 333 pc $cm^{-3}$ within each 1\,MHz channel.

%%%%% Data Analysis
%%%\subsection{Data Reduction}

%%%\subsubsection{single-pulse search}
We used the pulsar analysis software suite {\sc PRESTO}~\footnote{https://github.com/scottransom/presto} to process the Parkes search mode data. 
Previous observations show that radio emission from magnetar have very flat spectra.~\citep{kaspiMagnetars2017}. Therefore, the full 3.3\,GHz band width data sets were used to search for possible single pulses.  We also searched for possible limited band signals using data sub-banded into $704-1200$, $1200-1500$, $1500-2000$, $2000-2500$, $2500-3000$, $3000-3500$, $3500-4032$ MHz.
We used the routine {\sc rfifind} to identify strong narrow-band and short-duration broadband radio frequency interference (RFI) and produced RFI mask files. Our pipeline applied a 1.0\,s integration time for the RFI identification and a $6 \sigma$ cutoff to reject time-domain and frequency-domain interference. 
Our observation was coherently de-dispersed at the reported DM of 333 pc $cm^{-3}$. We searched DM trials in a range $\pm 10$ pc $cm^{-3}$ centered at the reported DM value with a DM step of 0.1 pc $cm^{-3}$. The {\sc prepdata} routine were then used to de-disperse the data at each of the trial DMs, and remove RFI based on the mask file.
Single pulse candidates with a signal-to-noise ratio (S/N) larger than seven were identified using the {\sc single\_pulse\_search.py} routine for each de-dispersed time series file and boxcar filtering with width up to 300 samples was used. All of the several thousands of candidates were grouped using the same method as described in \cite{zhang2020parkes}. For these groups, we only visually investigated the candidate with the highest S/N present within that group. 

%There are evidences that magnetar's radio pulsation can exist on a pretty wide band with a flat spectrum \citep{kaspiMagnetars2017}.
%So we directly search the whold 3.3GHz band width.
%We do single pulse using a widely-use software in pulsar field \emph{\sc PRESTO} 
%\footnote{https://github.com/scottransom/presto}.
%We first used command \emph{\sc rfifind} to identify radio frequency interference (RFI) and generate a mask to zap them out.
%Then \emph{\sc prepdata} dedisperse the data with a series of DM trial, which we set to be 0.1 per step and 20 range center at 333 pc $cm^{-3}$.
%It will form many one-dimension time-series ready for searching.
%And the \emph{\sc single\_pulse\_search.py} searched significant peaks with a series of boxcar filters of different width from 1 to 300 time samples.
%THe signal to noise ratio (SNR) was set to 7.
%The candidate wound then be selected according to their SNR and time of arrive.

%We also do a sub-band search to avoid narrow band RFI.
%we transformed the data into several sub-band first($700-1200, 1200-1500, 1500-2000, 2000-2500, 2500-3000, 3000-3500, 3500-4000$ MHz).
%Then we do the exact same searching on these sub-band data.

%%%\subsubsection{periodic search}
We searched for possible periodic signals using a similar manner to the single pulse searches. Both the full bandwidth and sub-banding data sets were processed.
RFI was rejected and marked using {\sc rfifind} and the DM trials are in a range $\pm 10$ pc $cm^{-3}$ centered at the 333 pc $cm^{-3}$ with a DM step of 0.1 pc $cm^{-3}$. As the latest spin period for SGR~1935+2154 in 2020 October was reported by FAST to be 3.24781s \citep{2020ATel14084....1Z}, we folded our observation using this period value at each trial DM using {\sc prepfold} routine.

%The spin period for SGR 1935+2154 in 2020 was measured by NICER's x-ray observation to be 3.24731 s \citep{younesNICERView20202020}. We therefore folded our observation 

%So instead of using blind-search, we directly folded the data into this period.

%We used \emph{\sc prepfold} command in \emph{\sc PRESTO} to do the folding.
%Before that, We added the data together to form a 1.6h file.
%We also added sub-band data and folded them separately.
%Then using \emph{\sc psrfits\_subband} to compress the data by combining 4 polarization.
%There are no any periodic signal been found in both data sets.
%We tried the period reported by MNC and FAST too, but also no finding.

%%%%% p4-Discussion
\section{Results}
\label{res}
36 single pulse candidates with S/N $\ge 7$ were detected. However, all of them were clearly caused by RFI and no convincing pulse from SGR~1935+2154 was detected. We also did not detect any convincing candidate from the periodicity-search. 

%single pulses
%%%results
%whole band 
%... candidates had been found.
%After the manual inspection, there is no real bursting single within them.

%narrow bands 
%We got 36 candidates, but there is no real burst among them.
%There are no any periodic signal been found in both data sets.

%\subsubsection{limitation }
Limits on the flux density of a radio pulse can be estimated as:
%The sensitivity limitation for a radio observation to a bursting signal can be described as follow.
\begin{equation}
S_{lim}=\frac{\sigma\,{\rm S/N}_{\rm min}\,T_{\rm sys}}{G \sqrt{{\Delta}{\nu}N_p{t}_{\rm obs}}}, 
\label{equ:limit}
\end{equation}
%\begin{equation}
%S_{\text {lim}}=\frac{\sigma S / N T_{\text {sys }}}{G \sqrt{\Delta \nu N_{p} t_{\text {obs }}}}
%\end{equation}
%\noindent 
%Where $S_{\rm peak}$ is the peak flux of a imaginary burst that can be detected with a relative S/N.
where a system temperature of T$_{\rm {sys}}=22$\,K, a loss factor $\sigma=1.5$ and telescope antenna gain $G=1.8$ for UWL receiver of Parkes telescope were used.  \citep{hobbsUltrawideBandwidth7042020}.
Assuming a pulse width of 0.5\,ms and flat spectrum, our non-detection of signal with S/N above 7 put a fluence limitation of 35 mJy\,ms for the full 3.3\,GHz bandwidth data sets. The limits of flux density and fluence of our single pulse search at different frequencies ranges are presented in Table~\ref{table:limits}.    

%put a 30mJy limitation to a radio burst within 704 to 4032MHz.
As for periodic signals, equation~\ref{equ:limit} should times $\sqrt{\frac{\delta}{1-\delta}}$ and $\delta$ is the duty cycle.
According to MNC detection \citep{2020ATel13783....1B}, we assume a pulse width of 100ms, corresponding to a duty cycle of 0.03. 
Our non-detection with the 1.6h observation of the full 3.3GHz band width put a 7$\sigma$ limit of 3.6$\mu$Jy. 
Limits of flux density and fluence of our periodicity search at different frequencies ranges are presented in Table~\ref{table:limits}.

\begin{table*}
\caption{Summary of the flux density and fluence limits of the single pulses and periodicity search of SGR~1935+2154 with Parkes UWL receiver}
\begin{center}
\begin{tabular}{ccccccccccl}
\hline
\hline
Freq. Range &  Assuming Width   & Assuming Width      &Flux Density Limit (7$\sigma)$          &Fluence Limit (7$\sigma)$  \\
   (MHz)     &  Single Pulse(ms) & Periodic Signal(ms) & Single Pulse/Periodic Signal  & Single Pulse/Periodic Signal    \\
\hline
$704-1200$  & 0.5 & 100 & $\rm 181mJy$ / $\rm 9.2\,\mu Jy$ & $\rm 91\,mJy\, ms$ / $\rm 0.92\,mJy\, ms$ &\\
$1200-1500$ & 0.5 & 100 & $\rm 234mJy$ / $\rm 11.9\,\mu Jy$ & $\rm 117\,mJy\, ms$ / $\rm 1.19\,mJy\, ms$ &\\
$1500-2000$ & 0.5 & 100 & $\rm 181mJy$ / $\rm 9.2\,\mu Jy$ & $\rm 91m\,Jy\, ms$ / $\rm 0.92\,mJy\, ms$ &\\
$2000-2500$ & 0.5 & 100 & $\rm 181mJy$ / $\rm 9.2\,\mu Jy$ & $\rm 91m\,Jy\, ms$ / $\rm 0.92\,mJy\, ms$ &\\
$2500-3000$ & 0.5 & 100 & $\rm 181mJy$ / $\rm 9.2\,\mu Jy$ & $\rm 91m\,Jy\, ms$ / $\rm 0.92\,mJy\, ms$ &\\
$3000-3500$ & 0.5 & 100 & $\rm 181mJy$ / $\rm 9.2\,\mu Jy$ & $\rm 91m\,Jy\, ms$ / $\rm 0.92\,mJy\, ms$ &\\
$3500-4032$ & 0.5 & 100 & $\rm 181mJy$ / $\rm 9.2\,\mu Jy$ & $\rm 91m\,Jy\, ms$ / $\rm 0.92\,mJy\, ms$ &\\
$704-4032$ & 0.5 & 100 & $\rm 70mJy$ / $\rm 3.6\,\mu Jy$ & $\rm 35m\,Jy\, ms$ / $\rm 0.36\,mJy\, ms$ &\\
\hline
\end{tabular}
\end{center}
\label{table:limits}
\end{table*}

\section{Discussion}
\label{dis}

%%% single pulses 
Our search of periodic signal and single pulses from SGR~1935+2154 with Parkes UWL receiver did not find any convincing signal. An integration of 1.6h observation allows us to derive 7$\sigma$ upper bounds on the fluence of $\rm 0.36\,mJy\,ms$ and $\rm 35\,mJy\,ms$ for the single pulse and periodicity search using the full 3.3\,GHz bandwidth, respectively. The single pulse fluence limit is slightly larger than the result of~\cite{Bailes2021} on April 2020 (i.e.  25\,mJy\,ms) and we 
noticed that~\citet{2020ATel14084....1Z} have carried out a one-hour observation of SGR~1935+2154 using FAST radio telescope just two days before our campaign.   
%Two days before our observation FAST have detected multiple bursts and periodic signal.
%\citep{2020ATel14084....1Z}.
The brightest single pulse detected by them has a fluence up to 40\,mJy\,ms, which is well above our fluence limit of the whole 3.3\,GHz band data sets, but below our limits using a bandwidth of 500\,MHz. Our results suggest that either the burst event rate of SGR1935 is reduced, or more likely, the spectrum of SGR1935 is not flat, or its single pulses are intrinsically narrow band. 

Our limit on the flux density of periodical signals using the full 3.3\,GHz bandwidth data sets is $\rm 3.6\,\mu Jy$, much lower than MNC's periodical detection of flux density of 4\,mJy on May 30 2020 \citep{2020ATel13783....1B} and CHIME's limit of 0.2\,mJy on May 30, 2020~\citep{2020ATel13838....1T}, and slightly lower than the Green bank telescope's limitation of $6.3 \rm \mu Jy$ on October 16, 2020~\citep{2020ATel14151....1S}. 
\citet{2020ATel14084....1Z} also claimed detection of periodic radio emission, however, no exact flux density or fluence measurement was presented. 
It is notable that the CHIME's limit of 0.2\,mJy was only 9 hours after the MNC's detection of 4\,mJy, which indicates a sharp of flux density of the periodic radio radiation. If the flux density of FAST detection is larger than our limit, this could be the second time that this phenomenon has been detected on SGR~1935+2154, which is similar to the intermittent pulsation behavior.  
%Intermittent pulsation behavior of magnetars
One of the six radio loud magnetars J1810-197 had shown intermittent pulsation behavior \citep{camiloRADIODISAPPEARANCEMAGNETAR2016}.
This source shut down radio pulsation in 2008 after a on state lasting 32 months.
It decreasing during the first 10 months but been steady for rest of the on period and suddenly wend off without any secular decrease.
%There is still a lack of theoretical explanation for such sharp radio turn-off.
%In some models it's believed that the radio emission mechanism are different between magnetars and ordinary radio pulsars %\citep{wangCoherentRadioEmission2019}.

However, if the flux density of FAST detection is much smaller than our limit, then it will show that maganitars could have periodic radiation with flux density that spans several orders of magnitude. The so-called ``shut down'' state of magnetars like J1810-197 could also be detected with weak emission in more sensitive observation.   
%Telescope with high sensitivity would more likely to reveal the mystery of magnetars.
Our limit of the periodic signal could derive that only telescopes with a diameter larger than 139\,m have chance to make a 10 $\sigma$ detection with one-hour observation with bandwidh of 300\,Mhz. Telescopes with high sensitivity like FAST are necessary to uncovering the radio activities for Magnetars like SGR~1935+2154.

\section*{Acknowledgements}
The Parkes radio telescope (``Murriyang'') is part of the Australia Telescope National Facility which is funded by the Australian Government for operation as a National Facility managed by CSIRO. This paper includes archived data obtained through the CSIRO Data Access Portal (\url{https://data.csiro.au}). This work was supported by ACAMAR Postdoctoral Fellow, the National Natural Science Foundation of China (Grant No. 11725314, 12041306, 11903019), China Postdoctoral Science Foundation (Grant No. 2020M681758).

%%%%%%%%%%%%%%%%%%%% REFERENCES %%%%%%%%%%%%%%%%%%

% The best way to enter references is to use BibTeX:

%\bibliographystyle{mnras}

\bibliographystyle{IEEEtran}

%\bibliography{Bib/Zotero, Bib/hm}
%\bibliography{hand made}

%%%%%%%%%%%%%%%%% APPENDICES %%%%%%%%%%%%%%%%%%%%%

%\appendix

%\section{Some extra material}

%%%%%%%%%%%%%%%%%%%%%%%%%%%%%%%%%%%%%%%%%%%%%%%%%%

% Don't change these lines
%\bsp	% typesetting comment
\label{lastpage}
\end{document}